\documentclass[12pt]{article}
\usepackage[english]{babel}
\usepackage{amssymb,latexsym,amsmath}

\begin{document}

\title
{Englert-type solutions of $d=11$ supergravity}
\author
{E.K. Loginov\footnote{ek.loginov@mail.ru}\\
\it Ivanovo State University\\
\it Ermaka St. 39, Ivanovo, 153025, Russia}
\maketitle

\begin{abstract}
A family of geometries on $S^7$ arise as solutions of the classical equations of motion in 11
dimensions. In addition to the conventional riemannian geometry and the two exceptional
Cartan-Schouten compact flat geometries with torsion, one can also obtain non-flat geometries
with torsion. This torsion is given locally by the structure constants of a nonassociative
geodesic loop in the affinely connected space.
\end{abstract}

\section{Introduction}

It is well known~\cite{duff86} that four-dimensional gravity, Yang-Mills interactions and
matter fields may originate from a higher-dimensional theory of pure gravity. This prospect
for unifying matter and the fundamental interactions including gravity is also offered by
supergravity. This is even more so because supersymmetry puts restrictions on the number of
possible space-time dimensions. The maximal dimension for which one can balance bosonic and
fermionic degrees of freedom with highest spin two is eleven. Supergravity in 11
dimensions~\cite{crem78} is thus possible to study spontaneous compactifications of this
theory, i.e. solutions of the 11-dimensional equations of motion for which the ground state
corresponds to a product space of a 4-dimensional space-time and a compact 7-dimensional
space.
\par
In the Bose sector of this theory the equations of motion (Einstein equations and equations
for the antisymmetric gauge field strength) have the form
\begin{align}
R_{MN}-\frac12g_{MN}R&=12\left(8F_{MPQR}F_{N}{}^{PQR}-g_{MN}F_{SPQR}F^{SPQR}\right),
\label{51-11}\\
F^{MNPQ}{}_{;M}&=-\frac{\sqrt2}{24}\varepsilon^{NPQM_1\dots
M_8}F_{M_1M_2M_3M_4}F_{M_5M_6M_7M_8}\label{51-12},
\end{align}
where $\varepsilon^{M_1\dots M_{r}}$ is a fully antisymmetric covariant constant tensor such
that $\varepsilon_{1\dots r}=|g|^{1/2}$. The Englert solution~\cite{engl82} compactifies
$d=11$ spacetime into a Riemann product $M=AdS_4\times S^7$ of two Einstein spaces, a
4-dimensional anti--de Sitter universe and the 7-sphere. The ansatz of Englert is to set
\begin{align}
F_{\mu\nu\sigma\lambda}&=\rho\varepsilon_{\mu\nu\sigma\lambda},\label{51-01}\\
F_{mnpq}&=\lambda\partial_{[q}S_{mnp]}\label{51-02},
\end{align}
where $\rho$ and $\lambda$ are real constants and $S_{mnp}=S_{[mnp]}$ is a suitable totally
antisymmetric torsion tensor. Note that the connection between an antisymmetric gauge field
strength and a torsion defined by (\ref{51-02}) has an universal character in the
11-dimensional supergravity. Bars and McDowell~\cite{bars83} have shown that the
$g_{MN}/A_{MNP}$ gravity-matter system may be reinterpreted, in first-order formalism, as a
pure gravity theory with torsion $S_{MNP}$ such that
\begin{align}
A_{MNP}&=\lambda S_{[MNP]},\label{51-22}\\
F_{MNPS}&=\partial_{[S}A_{MNP]}.\label{51-23}
\end{align}
In addition, in Ref.~\cite{logi09} was pointed out that the torsion tensor is fully
antisymmetric, i.e. $S_{MNP}=S_{[MNP]}$. In this sense the deformation
$\Gamma_{MNP}\to\Gamma_{MNP}+S_{MNP}$ of the Riemann connection converts $M$ into an affinely
connected space with torsion.
\par
In this paper we shall derive the surprising result that there exists not only three
Cartan-Schouten affine metric geometries on $S^7$, but a family of non-flat geometries on
$S^7$ which emerge as solutions of 11-dimensional supergravity. In Sec. 2, we construct an
one-parameter family of affine connections in the space of unit octonions. Then we find the
torsion and curvature tensors of the connections. In Sec. 3, we show that each torsion
satisfies the equations of motion of 11-dimensional supergravity.

\section{Extended Cartan-Schouten construction}

Although the notion of a space with an affine connections were initially created by Cartan and
Schouten independently, in 1926 the joint papers~\cite{cart26} and~\cite{cart26a} of both
geometers were published. Both papers were concerned with Riemannian geometry, but in both
cases, one way or another, the geometry of a space with an affine connection was discussed. In
the first of these papers, the authors considered three affine connections associated with any
Lie group. In the second paper, they considered the absolute parallelisms of the first and the
second type in the 7-sphere of the unite octonions. Later, the theory of affine connections
associated with any Lie group was developed by Akivis in his paper~\cite{akiv78}. In this
paper Akivis constructed, on the variety of any Lie group, an one-parameter affinely connected
family, containing the three Cartan-Schouten connections. In the section we apply the Akivis
method to construction of an one-parameter family of affine connections in $S^7$.
\par
Let $a$ be an arbitrary point of $S^7$. In a neighborhood of this point, we introduce a binary
operation as follows. Let $u$ and $v$ be two points belonging to this neighborhood. The
parallel displacement of vectors of the first type on $S^7$ may be determined by the right
translations $zR_{x}=zx$ in the 7-sphere of the unite octonions. Then the geodesic
$\overset\frown{au}$ is translated into the geodesic $\overset\frown{vw}$, where $v=aR_{x}$
and $w=uR_{x}$. Therefore
\begin{equation}\label{51-09}
w=u(a^{-1}v)\equiv u\circ v.
\end{equation}
This equation defines a geodesic loop $G_{a}$ with the unity $a$. Obviously, the loop $G_{a}$
is nonassociative and it is locally isotopic to the Moufang loop $S^7$. Recall that the set of
the unite octonions is closed relative to the multiplication in the algebra of octonions, and
therefore it is an analytic Moufang loop. On the other hand~\cite{bruc71,logi07}, every loop
isotopic to a Moufang loop is isomorphic to its principal isotope and every principal isotope
of $S^7$ is isomorphic to $S^7$. Therefore the loops $G_{a}$ and $S^7$ are locally isomorphic.
\par
Now we define an one-parameter family of loops $G_{a}^{\alpha}$ with the multiplication
\begin{equation}\label{54-01}
w=v^{\alpha}\circ u\circ v^{1-\alpha}\equiv u\ast v,
\end{equation}
where $\alpha$ is a real constant. It follows from the isomorphism of the loops $G_{a}$ that
the loops $G_{a}^{\alpha}$ are isomorphic to the loop $G^{\alpha}$ with the multiplication law
\begin{equation}\label{54-02}
w=v^{\alpha}uv^{1-\alpha}=(v^{\alpha}uv^{-\alpha})R_{v}.
\end{equation}
Let $v$ be a fixed point of $S^7$ and let the point $u$ runs its one-parameter subgroup $g$.
Then the point $u'=v^{\alpha}uv^{-\alpha}$ runs an one-parameter subgroup $g'$ and the point
$w$ runs a line obtained from $g'$ by the translation $R_{v}$. Hence if $a$ and $v$ are fixed
points of $S^7$ and the point $u$ describes the geodesic line $\overset\frown{au}$ generated
by a family of one-parameter subgroups, then the point $w$ describes the geodesic line
$\overset\frown{vw}$. Thus, all loops $G_{a}^{\alpha}$ are geodesic loops of affine
connections on $S^7$ which have the same system of geodesic lines generated by one-parameter
subgroups of the loop of the unite octonions.
\par
Now we shall find the torsion and curvature tensors of affine connections generated on $S^7$
by the geodesic loop $G_{a}^{\alpha}$. To this end we represent the multiplication
(\ref{54-02}) in the coordinate form. Since the loop $S^7$ is di-associative (i.e. any two
elements of $S^7$ generate an subgroup), it follows that in a neighborhood of the unity the
multiplication operation in $S^7$ is expressed through the addition and multiplication
operations in the tangent algebra by the usual Campbell-Hausdorff series
\begin{equation}
(xy)^{i}=x^{i}+y^{i}+\frac12\hat c^{i}_{jk}x^{j}x^{k} +\frac{1}{12}\hat c^{i}_{jm}\hat
c^{m}_{kl}(x^{j}x^{k}y^{l}+y^{j}y^{k}x^{l})+\dots,
\end{equation}
where $\hat c^{i}_{jk}$ are structure constants of the algebra of octonions. Twice using this
formula, we find coordinates of the point $w$ that defined by Eq.~(\ref{54-02}):
\begin{multline}\label{54-03}
w^{i}=u^{i}+v^{i}+\frac12(1-2\alpha)\hat c^{i}_{jk}u^{j}v^{k}\\
+\frac{1}{12}\hat c^{i}_{jm}\hat c^{m}_{kl}\left[u^{j}u^{k}v^{l}
+(1-6\alpha+6\alpha^2)v^{j}v^{k}u^{l}\right]+\dots.
\end{multline}
On the other hand, if we introduce local coordinates in a neighborhood of unity of
$G^{\alpha}$, then the multiplication in $S^7$ is expressed by
\begin{equation}\label{54-04}
w^{i}=u^{i}+v^{i}+\lambda^{i}_{jk}u^{j}v^{k}
+\frac12\left(\mu^{i}_{jkl}u^{j}u^{k}v^{l}+\nu^{i}_{jkl}u^{j}v^{k}v^{l}\right)+\dots.
\end{equation}
We define the tensors
\begin{align}
\alpha^{i}_{jk}&=\lambda^{i}_{[jk]},\label{54-05}\\
-2\beta^{i}_{jkl}&=\mu^{i}_{[jk]}-\nu^{i}_{[jk]}+\lambda^{m}_{jk}\lambda^{i}_{ml}
-\lambda^{i}_{jm}\lambda^{m}_{kl}.\label{54-06}
\end{align}
These tensors are called the fundamental tensors of the geodesic loop $G^{\alpha}$. Note that
these tensors are structure constants of the binary-ternary tangent algebra of the geodesic
loop~\cite{akiv76} (see also~\cite{akiv06}). Comparing (\ref{54-03}) and (\ref{54-04}), we get
\begin{align}
2\alpha^{i}_{jk}&=(1-2\alpha)\hat c^{i}_{jk},\label{54-14}\\
-4\beta^{i}_{jkl}&=\alpha(1-\alpha)\hat c^{i}_{jm}\hat c^{m}_{kl} +(1-3\alpha+3\alpha^2)\hat
c^{i}_{m[j}\hat c^{m}_{kl]}.
\end{align}
It follows from (\ref{54-14}) that the fundamental tensor $\alpha^{i}_{jk}$ is defined by
\begin{equation}\label{51-13}
\alpha^{i}_{jk}=k c^{i}_{jk},
\end{equation}
where $c^{i}_{jk}$ are standard structure constants of the algebra of octonions and $k$ is a
real number.
\par
It is known~\cite{akiv78} that for any geodesic loop constructed in a neighborhood of a point
$a$ of an affinely connected space, the fundamental tensors can be expressed using values of
the torsion and curvature tensors in $a$ by the formulas
\begin{align}
\alpha^{i}_{jk}&=-S^{i}_{jk},\label{51-07}\\
4\beta^{i}_{jkl}&=-2\nabla_{l}S^{i}_{jk}-R^{i}_{jkl}.\label{51-08}
\end{align}
Note that we define all these tensors, as is done in ref.~\cite{logi09}. Since the tensor
$c_{ijk}$ is fully antisymmetric, it follows from (\ref{51-13}) and (\ref{51-07}) that the
geodesic loops $G_{a}^{\alpha}$ (the index $\alpha$ is fixed) generate the metric-compatible
affine connection on $S^7$:
\begin{equation}\label{54-15}
\Gamma_{ijk}=\overset\circ\Gamma_{ijk}+S_{ijk},
\end{equation}
where $\overset\circ\Gamma_{ijk}$ is the riemannian symmetric connection and $S_{ijk}$ is a
fully antisymmetric torsion. Using the full skew-symmetry of $S_{ijk}$, we can rewrite the
tensor (\ref{51-08}) in the form
\begin{equation}\label{54-16}
4\beta_{ijkl}=-2S_{ijk;l}+6S^{m}_{[ij}S_{kl]m}-R_{ijkl}.
\end{equation}
Let $\alpha=0$. Then we have the remaining curvature-less Cartan-Schouten geometry of absolute
parallelism on $S^7$. Using the cyclic identities for $R_{ijkl}$ one easily obtains the
conditions for such torsion, and hence for the parallelism:
\begin{equation}\label{54-13}
S_{ijk;l}=S^{m}_{[ij}S_{kl]m}.
\end{equation}
Now we consider the case of arbitrary $\alpha\ne\frac12$. The case $\alpha=\frac12$
corresponds to the torsionless riemannian geometry, and therefore it is not interesting. Using
Eqs. (\ref{54-14}) and (\ref{51-07}), we get
\begin{equation}\label{54-17}
S_{ijk;l}=hS^{m}_{[ij}S_{kl]m},\qquad h=\frac{1}{1-2\alpha}
\end{equation}
instead of (\ref{54-13}). Using Eqs. (\ref{54-16}) and (\ref{54-17}), we obtain the curvature
tensor of affine connection on $S^7$:
\begin{equation}\label{54-10}
R_{ijkl}=4\alpha(1-\alpha)S^{m}_{ij}S_{klm}-4\alpha(2-3\alpha)S^{m}_{[ij}S_{kl]m}.
\end{equation}
Obviously, the curvature tensor is equal to zero if $\alpha=0$ and it is fully antisymmetric
if $\alpha=1$. The corresponding geodesic loops $G_{a}^0$ and $G_{a}^1$ are locally isomorphic
to the Moufang loop $S^7$. Solutions of the classical equations of motion of 11-dimensional
supergravity connected with these loops was found in Refs.~\cite{engl82,logi09}. In the next
section we shall construct solutions of 11-dimensional supergravity connected with nonmoufang
geodesic loops.

\section{The Englert-type solutions}

We consider the Bose sector of $d=11$ supergravity as a pure gravity theory with torsion and
suppose that the matter fields have non-vanishing components in the internal space $S^7$.
Substituting the Freund-Rubin ansatz (\ref{51-01}) in Eq. (\ref{51-12}), we obtain
\begin{equation}\label{51-27}
F^{mnpq}{}_{;m}=\sqrt2\rho\varepsilon^{npqijkl}F_{ijkl},
\end{equation}
where $\varepsilon^{npqijkl}$ is the fully antisymmetric covariant constant 7-tensor. It
follows from (\ref{51-22}) and (\ref{51-23}) that solutions of Eq. (\ref{51-27}) are related
to torsion tensors by the relation (\ref{51-02}). We shall therefore first derive analogs of
the Cartan-Schouten-Englert equations~\cite{engl82,cart26a} which these tension tensors must
satisfy. We shall need to the following seven-dimensional algebraic
identities~\cite{logi09,dund84}:
\begin{align}
\alpha_{ijm}\alpha^{ijn}&=6k^2\delta^{n}_{m},\label{51-29}\\
\alpha_{im}{}^{j}\alpha_{jn}{}^{k}\alpha_{kp}{}^{i}&=3k^2\alpha_{mnp},\label{51-30}\\
\beta_{mijk}\beta^{nijk}&=24k^4\delta^{n}_{m},\label{51-52}\\
\varepsilon^{npqlijk}\beta_{ijkl}&=24k\alpha^{npq}.\label{51-32}
\end{align}
Using the identities  (\ref{54-17}), (\ref{51-29}), and (\ref{51-30}), we get
\begin{equation}
S_{npq;m}{}^{;m}+(2hk)^2S_{npq}=0.
\end{equation}
By taking into account the obtained identities, we rewrite the equations (\ref{51-27}) as
\begin{equation}\label{51-37}
(2hk)^2S^{npq}+\sqrt2\rho\varepsilon^{npqijkl}S_{ijk;l}=0.
\end{equation}
Substituting the ansatz (\ref{54-17}) in (\ref{51-37}) and using the relations of self-duality
(\ref{51-32}), we find the condition
\begin{equation}\label{54-11}
h=6\sqrt2\rho k^{-1},
\end{equation}
in which Eq. (\ref{51-37}) turn into an identity. Thus, if we choose the gauge field strength
$F_{mnpq}$ in the form (\ref{51-02}), then we get solutions of the equations (\ref{51-27}).
\par
Now we consider the Einstein equations (\ref{51-11}). It is obvious that these equations can
be satisfied if $F_{mrpq}F_{n}{}^{rpq}$ is proportional to $g_{mn}$. This is indeed the case
as follows from Eqs. (\ref{54-17}) and (\ref{51-52}):
\begin{align}
F_{\mu\sigma\rho\lambda}F_{\nu}{}^{\sigma\rho\lambda}&=-6\rho^2g_{\mu\nu},\\
F_{mrpq}F_{n}{}^{rpq}&=24k^4h^2\lambda^2g_{mn}.
\end{align}
Substituting these relations in the Einstein equations (\ref{51-11}), we get
\begin{align}
\overset\circ {R}_{mn}&=6(hk)^2g_{mn},\\
\overset\circ {R}_{\mu\nu}&=-10(hk)^2g_{\mu\nu},
\end{align}
where the constant
\begin{equation}
2\lambda^2=(12k)^{-2}.\label{54-12}
\end{equation}
It follows from Eqs. (\ref{54-11}) and (\ref{54-12}) that the constants $k$ and $h$ are
determined by $\rho$ and $\lambda$ which still arbitrary. Thus we have obtained, in addition
to the solutions with $h=\pm1$ that was found in Refs.~\cite{engl82,logi09}, new solutions
where the four-scalars take the values (\ref{51-01}) and (\ref{51-02}) with arbitrary $\rho$
and $\lambda$.
\par
Note that in the paper~\cite{engl82} Englert constructed two independent solutions
$S^{+}_{mnp}$ and $S^{-}_{mnp}$ of the equations of motion. These solutions are connected with
two types of parallelisms on $S^7$. The first of them coincides precisely with torsion
generated on $S^7$ by the geodesics loops $G^0_{a}$. In order that to obtain the second, we
suppose $h=(2\alpha-1)^{-1}$ in Eq. (\ref{54-17}). Then $S^{-}_{mnp}$ will be coincided with
torsion generated on $S^7$ by the loops $G^1_{a}$. By repeating the previous reasoning, we get
a new series of solutions of 11-dimensional supergravity.

\end{document}